\documentclass[conference]{IEEEtran}
\IEEEoverridecommandlockouts

\usepackage{epsfig,rotating,setspace,latexsym,amsmath,epsf,amssymb,amsfonts,bm,theorem,cite,algorithm,graphicx,epsf,authblk,epstopdf,color,algpseudocode,bbm,dsfont}

\allowdisplaybreaks

\begin{document}

\title{Remote Short Blocklength Process Monitoring: Trade-off Between Resolution and Data Freshness}


\author{Stefan Roth$^*$, Ahmed Arafa$^\S$, H. Vincent Poor$^\ddagger$, Aydin Sezgin$^*$ \\
$^*$Institute of Digital Communication Systems, Ruhr University Bochum, Germany\\
$^\S$Department of Electrical and Computer Engineering, University of North Carolina at Charlotte, NC, USA\\
$^\ddagger$Electrical Engineering Department, Princeton University, NJ, USA\\
Email: \{stefan.roth-k21, aydin.sezgin\}@rub.de, aarafa@uncc.edu, poor@princeton.edu
\thanks{This work was supported in part by the U.S. National Science Foundation under Grant CCF-1908308, and in part by the Deutsche Forschungsgemeinschaft (DFG, German Research Foundation) under Germany's Excellence Strategy - EXC 2092 CASA - 390781972.}
}


\maketitle

\begin{abstract}
In cyber-physical systems, as in 5G and beyond, multiple physical processes require timely online monitoring at a remote device. There, the received information is used to estimate current and future process values. When transmitting the process data over a communication channel, source-channel coding is used in order to reduce data errors. During transmission, a high data resolution is helpful to capture the value of the process variables precisely. However, this typically comes with long transmission delays reducing the utilizability of the data, since the estimation quality gets reduced over time. In this paper, the trade-off between having recent data and precise measurements is captured for a Gauss-Markov process. An Age-of-Information (AoI) metric is used to assess data timeliness, while mean square error (MSE) is used to assess the precision of the predicted process values. AoI appears inherently within the MSE expressions, yet it can be relatively easier to optimize. Our goal is to minimize a time-averaged version of both metrics. We follow a {\it short blocklength} source-channel coding approach, and optimize the parameters of the codes being used in order to describe an {\it achievability region} between MSE and AoI.
\end{abstract}

\IEEEpeerreviewmaketitle

\section{Introduction}

Traditionally, communication systems have been considered for asymptotically large blocklengths. However, in delay-sensitive applications conducted in 5G and beyond, such as internet-of-things and networked control, it is critical to transmit data in short packets. In these cases, the asymptotic capacity cannot be achieved due to the impacts of channel and source dispersions \cite{5452208, 6408177}. Channel coding rates when using short blocklengths have been investigated in~\cite{5452208}. As some transmissions require both source and channel coding with short blocklengths, relations between communication rates and distortion effects have been jointly investigated in \cite{6408177}. In this paper, a short blocklength source-channel coding technique is applied to a time-varying physical process that needs to be monitored and estimated online at a remote location. To evaluate the performance, we use a combination of the \textit{Age-of-Information (AoI)} metric \cite{6195689} to assess timeliness of the estimates, and the \textit{mean square error (MSE)} to assess the accuracy of the estimates. AoI is defined as the time elapsed since the generation time of the latest data measurement that has been received.

Maintaining low AoI at remote devices (receivers) is useful to diagnose errors and detect anomalies within the data quickly, such that immediate action can be taken. Compared to MSE, AoI is relevant in situations for which the system dynamics are hard to keep track of. MSE can also be shown equal to (a function of) AoI in situations for which the process values are non-observable \cite{8006542, DBLP:journals/corr/abs-1902-03552}. In general, as we show in this paper, the two metrics are closely intertwined.

We focus on a Gauss-Markov process structure in the physical system. As the transmission is done in packets, the measurement data available at the receiver might become partially outdated during and after the transmission. However, the information received from the measurement also remains partially valuable and can be used to predict, in real-time, later process values until the arrival of a new measurement. This means that with an increasing AoI, the estimator becomes less observant of the process, and the MSE increases as well. Besides, the data received  is distorted by noise. When optimizing all this, a more precise quantization leads to a larger channel blocklength and thus, a higher transmission delay. 
Hence, the MSE of the estimate depends on the amount of data transmitted in each packet in two ways: $(1)$ when the amount of information is large, very accurate data can be transmitted, but this also comes with longer transmission delays; $(2)$ when the amount of information is small, transmission delays become short, whereas the data stored in a packet might be inaccurate. In order to optimize the time-averaged MSE of the estimation of the physical process at the receiver, an intermediate packet size is, therefore, expected to be optimal.

\begin{figure*}
	\centering
	\includegraphics[]{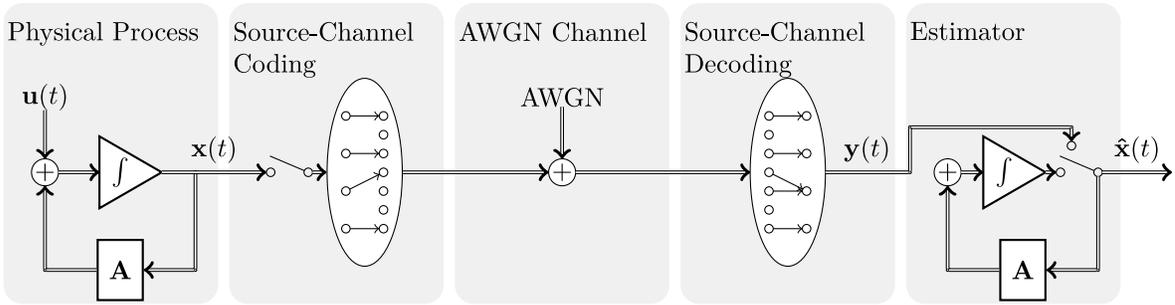}
	\caption{Schematic description of the different parts of the system.}
	\label{fig:Schematic}
\end{figure*}

Since both the MSE and AoI are time-varying, we jointly optimize a time-averaged version of both metrics in a {\it short blocklength source-channel coding} framework. The goal is to find the optimal channel blocklength (and hence transmission time), while engineering the values of the probability of transmission error and the tolerated distortion in the received data. {\it We show that there exists an achievable region describing a tradeoff between MSE and AoI,} which provides further insights on the intrinsic relationship between both parameters in real-time remote monitoring systems.

\subsection{Related Works}

Several works in the literature consider coding for AoI improvement, e.g., \cite{8006504, 8445909, 8437671, 8849713, 7925903, 8006541, 8377368, DBLP:journals/corr/abs-1905-03238}, of which \cite{8437671} considers a study of short channel blocklengths in an AoI/delay minimization framework. Other works focus on estimation frameworks with AoI considerations, e.g., \cite{8006542, DBLP:journals/corr/abs-1902-03552, DBLP:journals/corr/abs-1903-12472, 8761710, 8619768, 8814627, 7282683}. The notion of updates with distortion has been studied in \cite{8988940}. Our work is different in the sense that we consider a joint source-channel short blocklength coding framework to describe the relationship between MSE and AoI.

\subsection{Notation}
Matrices (vectors) are denoted in uppercase $\mathbf{A}$ (lowercase $\mathbf{a}$). $\mathbf{I}_k$ refers to the identity matrix of size $k\times k$, and $e^{\mathbf{A}t}$ represents the matrix exponential.

\section{System Model}

The system consists of the following parts (see Fig.~\ref{fig:Schematic}):
\begin{itemize}
	\item A \textit{physical process} generating Gaussian distributed variables that are varying over time.
	\item These variables are encoded via \textit{joint source-channel coding} such that they can be recovered after being transmitted through the channel.
	\item The \textit{channel} is assumed to be an additive white Gaussian noise (AWGN) channel. This limits the probability of successful decoding. Besides, a transmission delay is added as well.
	\item At the receiver, \textit{joint source-channel decoding} is used to decode the values of the variables.
	\item The decoded variables are fed to an \textit{estimator} to predict the values of the following time instances until a new variable is decoded.
\end{itemize}
We elaborate on these components over the next subsections.

\subsection{The Physical Process}

We consider a linear physical process that can be described by a state-space equation. In case the process is non-linear, it is often possible to create a linear approximation nearby an operating point, such that the same methods are applicable \cite{7076051}. Our system model evolves similarly to a form in \cite{Astrom04feedbacksystems}:
\begin{subequations}\label{eq:stateSpaceEquation}
\begin{align}
\mathbf{\dot{x}}(t)&=\mathbf{A}\mathbf{x}(t)+\mathbf{u}(t),\label{eq:stateSpaceEquation1}\\
\mathbf{x}(0)&=\mathbf{x}_0,
\end{align}
\end{subequations}
where $\mathbf{u}(t)\sim\mathcal{N}(\mathbf{0},\mathbf{Q}_u)$ is the input noise, which is independent and identically distributed (i.i.d.) for all time instances $t$. 
The state of the system is described by the vector $\mathbf{x}(t)$ of dimension $k\times 1$. The initial state value at time $0$ is given by $\mathbf{x}(0)=\mathbf{x}_0$ and is bounded.
From this, the state at time $t$ can be described explicitly by the equation of motion given as the following solution of \eqref{eq:stateSpaceEquation}:
\begin{align}
\mathbf{x}(t)&=e^{\mathbf{A}t}\mathbf{x}_0+\int_{0}^{t}e^{\mathbf{A}(t-\mu)}\mathbf{u}(\mu)d\mu,&t\geq0.\label{eg:motionX}
\end{align}


\subsection{Data Freshness and Estimators}

The system bases on a Gauss-Markov process, which contains a state value that is changing over time. Hence, it is important for the system performance to have fresh data available at the receiver. The AoI metric is well-suited to capture this notion. If the latest measurement available at the receiver at time $t$ has been generated at time $\nu_t$, then the AoI is given by
\begin{align}
\tau=t-\nu_t,\quad t\geq\nu_t.\label{eq:AoI}
\end{align}
Note that this variable does not depend on the data resolution. It will be shown later that optimizing the AoI is not always equivalent to optimizing the measurement accuracy.

In \eqref{eg:motionX}, the system state is shown to depend on $\mathbf{x}_0$. When utilizing the sample value $\mathbf{x}(\nu_t)$ instead, this becomes
\begin{align}
\mathbf{x}(t)&=e^{\mathbf{A}\tau}\mathbf{x}(\nu_t)+\int_{0}^{\tau}e^{\mathbf{A}(\tau-\mu)}\mathbf{u}(\mu+\nu_t)d\mu,& t\geq\nu_t.\label{eg:motionXnu}
\end{align}
The sample value $\mathbf{x}(\nu_t)$ is transmitted and the decoded version at the receiver is referred to as $\mathbf{y}(\nu_t)$, based on which an estimate $\mathbf{\hat{x}}(t)$ of the state is formed, which we later specify. We will see that as the AoI $\tau$ increases, the difference between the system state and the estimate increases as well, which shows that the accuracy of this estimate is decreasing as long as no new measurement arrives. The MSE describing the difference between system state and the estimate is given by
\begin{align}
M(t)&=\mathds{E}\left[\left\|\mathbf{x}(t)-\mathbf{\hat{x}}(t)\right\|_2^2\right]\label{eq:MSEdef},
\end{align}
where $\|\cdot\|_2$ denotes the Euclidean norm.

For the case of no transmission errors, i.e., $\mathbf{y}(\nu_t)=\mathbf{x}(\nu_t)$, one can show that the optimal estimate is $\mathbf{\hat{x}}(t)=e^{\mathbf{A}\tau}\mathbf{y}(\nu_t)$, and therefore the MSE only describes the impacts of the transmission delay and is denoted in this case as $M^{\mathrm{D}}(t)$.
When employing $\mathbf{x}(t)$ as described in \eqref{eg:motionXnu}, this becomes
\begin{align}
M^{\mathrm{D}}(t)&=\mathds{E}\left[\bigg\|e^{\mathbf{A}\tau}\mathbf{x}(\nu_t)+\int_{0}^{\tau}e^{\mathbf{A}(\tau-\mu)}\mathbf{u}(\mu+\nu_t)d\mu\right.\nonumber\\&\hspace{4.5cm}\left.-e^{\mathbf{A}\tau}\mathbf{y}(\nu_t)\bigg\|_2^2\right]\\
&=\mathds{E}\left[\left\|\int_{0}^{\tau}e^{\mathbf{A}(\tau-\mu)}\mathbf{u}(\mu+\nu_t)d\mu\right\|_2^2\right].
\end{align}
As the input noise values $\mathbf{u}(t)$ and $\mathbf{u}(\mu)$ are uncorrelated for each $t\neq\mu$, this can be rephrased as
\begin{align}
M^{\mathrm{D}}(t)
&=\mathrm{trace}\left\{\int_{0}^{\tau}e^{\mathbf{A}(\tau-\mu)}\mathbf{Q}_ue^{\mathbf{A}^H(\tau-\mu)}d\mu\right\}.\label{eq:MnoChannelNoiseSolution}
\end{align}
Thus, a large AoI $\tau$ will also lead to a large MSE $M^{\mathrm{D}}(t)$ in case of no transmission errors.

\subsection{Short Blocklength Source-Channel Coding}

In practical systems, each data packet contains quantized and encoded information about the vector $\mathbf{x}(t)$ using a finite blocklength source-channel coding scheme. This introduces decoding errors and distortion with a non-zero probability.

Following the framework in \cite{6408177}, we aim at designing a system in such a way that the distortion exceeds a certain tolerable value $d$ with a pre-specified probability $\varepsilon$. In particular, for such condition to be satisfied, $k$ source symbols encoded into a channel code of length $n$ should satisfy \cite{6408177}
\begin{align}
	nC-kR(d)&\approx\sqrt{nV_C+kV_S}Q^{-1}(\varepsilon),\label{eq:shortBlocklengthAWGN}
\end{align}
where $Q(x)=\int_{x}^{\infty}\frac{1}{\sqrt{2\pi}}\exp\left(-\frac{y^2}{2}\right)dy$; $C$ is the channel capacity and $R(d)$ is the rate-distortion function \cite{Cover:2006:EIT:1146355}; $V_C$ is the channel dispersion \cite{5452208}; and $V_S$ is the source dispersion \cite{6145679}. Computing the above for our Gaussian source setting, communicated over an AWGN channel with SNR $P$, we have the capacity given by \cite{Cover:2006:EIT:1146355}
\begin{align}
	C=\frac{1}{2}\log_2\left(1+P\right).
\end{align}
Moreover, the rate-distortion function $R(d)$, which represents the number of bits needed to represent the source symbol with distortion not surpassing $d$, is given by 
\cite{e20060399,Cover:2006:EIT:1146355}
\begin{align}
R(d)&=\frac{1}{k}\sum_{i=1}^{k}\max\left\{\frac{1}{2}\log_2\left(\frac{\lambda_i\{\mathbf{Q}_y\}}{d}\right),0\right\},\label{eq:rateDistortion}
\end{align}
where $\mathbf{Q}_y$ denotes the covariance matrix of the output of the receiver, and $\lambda_i\{\mathbf{Q}_y\}$ denotes its $i$th eigenvalue. Finally, we have \cite{5452208, 6145679}
\begin{align}
	V_C&=\frac{1}{2}\left(1-\frac{1}{(1+P)^2}\right)\log_2^2(e), \\
	V_S&=\frac{1}{2}\log_2^2(e).
\end{align}

For a given set of system parameters, i.e., $d$, $\varepsilon$, and $P$, \eqref{eq:shortBlocklengthAWGN} provides a relation between source and channel blocklengths such that the distortion surpasses $d$ only $\varepsilon$ portion of the time. From this expression it follows that
\begin{align}
n^2C^2-2knCR(d)+k^2R^2(d)\approx\left(nV_C+kV_S\right)\left(Q^{-1}(\varepsilon)\right)^2,
\end{align}
from which we further have
\begin{align}
	n&=\frac{V_C\left(Q^{-1}(\varepsilon)\right)^2+2kCR(d)+\sqrt{\Delta}}{2C^2},\label{eq:nFromShortBlocklength}
\end{align}
with
\begin{align}
	\Delta&=\left(V_C\left(Q^{-1}(\varepsilon)\right)^2+2kCR(d)\right)^2\nonumber\\&\hspace{2.5cm}-4C^2\left(k^2R^2(d)-kV_S\left(Q^{-1}(\varepsilon)\right)^2\right)\nonumber\\
	&=V_C^2\left(Q^{-1}(\varepsilon)\right)^4+4k\left(V_CCR(d)+V_SC^2\right)\left(Q^{-1}(\varepsilon)\right)^2.\nonumber
\end{align}

Whenever the distortion surpasses $d$, a NACK is sent back to the transmitter, and a {\it new} measurement is acquired and transmitted. Otherwise, an ACK is sent back. It now follows that the probability of a NACK is $\varepsilon$.\footnote{We are assuming that the receiver is capable of verifying high distortion measurements via checking mechanisms that we do no explicitly discuss in this paper.}

\subsection{Sampling and Transmission Delay}

For a given channel blocklength $n$, we model the time incurred to traverse through the channel $r$ by the following linear model:
\begin{align}
r=\alpha n+\beta,\label{eq:lengthToTransmissionTime}
\end{align}
where $\alpha$ represents the symbol duration and $\beta$ refers to an extra channel-induced delay.

Now since each data packet is successfully decoded within the tolerable distortion with probability $1-\varepsilon$, it follows that the waiting time in between two consecutive successful receptions is $r^\prime=(m+1)r$, where $m$ denotes the number of failures, which is geometrically distributed with parameter $\varepsilon$. Thus, $r^\prime$ is distributed as follows
\begin{align}
r'\sim\sum_{m=0}^{\infty}\varepsilon^m(1-\varepsilon)\delta\left(r'-(m+1)r\right),\label{eq:yPrimeDistribution}
\end{align}
where $\delta(r^\prime)$ refers to the Dirac impulse.

We follow a fixed-waiting sampling policy, in which a new sample is acquired/transmitted following an ACK after $s$ time units. Such $s$ is the smallest possible value allowed by the system being considered, since the AoI is always equal to a fixed value $r$ following successful transmission.

\section{Problem Formulation}

Let $\mathbf{w}(t)$ denote the remaining distortion within a successfully received data packet. Hence, the received signal at the receiver is given as
\begin{align}
\mathbf{y}(\nu_t)&=\mathbf{x}(\nu_t)+\mathbf{w}(\nu_t).\label{eq:stateSpaceEquation2}
\end{align}
The distortion $\mathbf{w}(t)$ is zero-mean and is approximated as an additive Gaussian noise with covariance matrix $\mathbf{Q}_w=q_w\mathbf{I}_k$, in which $q_w$ is no larger than $d$. At a later point of time following successful reception, the system-state can be estimated using an MSE-optimal estimator $\mathbf{F}_{\tau}$ given by
\begin{align}
\mathbf{\hat{x}}(t)=\mathbf{F}_{\tau}\mathbf{y}(\nu_t),\quad t\geq\nu_t.\label{eq:mseestimator}
\end{align}

Let us now denote by $M^{\mathrm{C}}(t)$ the MSE in case of having no system-input $\mathbf{u}(t)$ after a packet has been transmitted, i.e., one that describes only the impacts of the channel noise and distortion. Using \eqref{eg:motionXnu}, \eqref{eq:stateSpaceEquation2}, and \eqref{eq:mseestimator}, this is given by
\begin{align}
&M^{\mathrm{C}}(t)=\mathds{E}\left[\left\|e^{\mathbf{A}\tau}\mathbf{x}(\nu_t)-\mathbf{F}_{\tau}\left(\mathbf{x}(\nu_t)+\mathbf{w}(\nu_t)\right)\right\|_2^2\right]\nonumber\\
&\hspace{0.5cm}=\mathrm{trace}\left\{\left(e^{\mathbf{A}\tau}-\mathbf{F}_{\tau}\right)\mathbf{Q}_x\left(e^{\mathbf{A}^H\tau}-\mathbf{F}_{\tau}^H\right)+\mathbf{F}_{\tau}\mathbf{Q}_w\mathbf{F}_{\tau}^H\right\},\label{eq:MnoDelay}
\end{align}
where $\mathbf{Q}_x=\lim\limits_{t\rightarrow\infty}\mathds{E}\left[\mathbf{x}(t)\mathbf{x}^H(t)\right]$ is the covariance of \eqref{eg:motionX} achieved at steady state, i.e., for large $t$.
Taking derivative of the above and setting it to $0$, the optimal estimator can be obtained as
\begin{align}
\mathbf{F}_{\tau}=e^{\mathbf{A}\tau}\mathbf{Q}_x\left(\mathbf{Q}_x+\mathbf{Q}_w\right)^{-1}.
\end{align}

In the following, the impacts of data freshness and channel noise are combined and joint expressions for MSE and AoI are created.

\subsection{Mean Square Error}
Recall that, the MSE has been described in \eqref{eq:MnoChannelNoiseSolution} for the idealized case of distortion-free transmission and in \eqref{eq:MnoDelay} for the idealized case of not having any input noise. Whereas the former depends on the input noise $\mathbf{u}(t)$, the latter depends on the channel noise and distortion $\mathbf{w}(t)$. As these two variables are uncorrelated, the time-varying MSE can be phrased as
\begin{align}
M(t)=M^{\mathrm{D}}(t)+M^{\mathrm{C}}(t).\label{eq:MgeneralSolution}
\end{align}
This value describes the mean square error of the estimate at a given point of time $t$. With an increasing AoI $\tau$, the first part increases leading to a higher MSE.

As in \cite{8000687}, the long term average MSE is formulated as
\begin{align}
\mathrm{MSE}=\limsup\limits_{n\rightarrow\infty}\frac{\mathds{E}\left[\int_{0}^{D_n}M(t)dt\right]}{\mathds{E}\left[D_n\right]}\label{eq:limMInfinite},
\end{align}
where $D_n$ is the reception time of the $n$th successful transmission. This reduces to minimizing the MSE over each successful transmission, since the transmission policy is stationary. That is, the numerator becomes 

\begin{align}
L(r,s,r')=\int_{r}^{r+s+r'}M(t)d\tau,\label{eq:mseIntegration}
\end{align}
with the denominator given by $\mathds{E}\left[s+r'\right]$. Therefore, \eqref{eq:limMInfinite} can be reformulated as \cite{8000687}
\begin{align}
\mathrm{MSE}=\frac{\mathds{E}\left[L(r,s,r')\right]}{\mathds{E}\left[s+r'\right]}.\label{eq:limMFinite}
\end{align}
We will also be interested in studying the two individual components constituting $M(t)$, i.e., $M^D(t)$ and $M^C(t)$, and will refer to their long term time-averages by $\mathrm{MSE}^D$ and $\mathrm{MSE}^C$, respectively. Therefore, we it holds that $\mathrm{MSE}=\mathrm{MSE}^D+\mathrm{MSE}^C$.

\subsection{Age of Information}
The AoI has been described fully in \eqref{eq:AoI}. Similar to the MSE in \eqref{eq:limMInfinite}, the time-average AoI can be expressed as
\begin{align}
\mathrm{AoI}=\limsup\limits_{n\rightarrow\infty}\frac{\mathds{E}\left[\int_{0}^{D_n}t-\nu_t dt\right]}{\mathds{E}\left[D_n\right]}.\label{eq:limAoiInfinite}
\end{align}
Similar to the MSE case, under a stationary sampling policy, the above can be transformed into a minimization over each successful transmission, i.e., the numerator becomes
\begin{align}
\int_{r}^{r+s+r'}\tau d\tau =\frac{1}{2}\left(\left(r+s+r'\right)^2-r^2\right).\label{eq:aoiIntegration}
\end{align}
Therefore, \eqref{eq:limAoiInfinite} can be expressed as
\begin{align}
\mathrm{AoI}&=\frac{\mathds{E}\left[(s+r')r+\frac{1}{2}(s+r')^2\right]}{\mathds{E}\left[s+r'\right]}.\label{eq:aoiAvg}
\end{align}

\subsection{Multi-Objective Optimization}

The goal is to optimize the parameters $d$ and $\varepsilon$ such that the weighted sum of MSE and AoI is minimized. Observe that for a given set of $d$ and $\varepsilon$ (and the system's parameter $k$) $n$ is given by \eqref{eq:nFromShortBlocklength}. To characterize the Pareto-boundary, the MSE and AoI have to be optimized jointly as follows:
\begin{subequations}\label{eq:JointOptProblem}
	\begin{align}
	\underset{d,\varepsilon}{\mathrm{\ minimize\ }}  & \begin{pmatrix}
	\mathrm{MSE},& \mathrm{AoI}
	\end{pmatrix}\tag{\ref{eq:JointOptProblem}}\\
	\mathrm{\ subject\ to\ }
	&~ \eqref{eq:nFromShortBlocklength},\ 
	\eqref{eq:lengthToTransmissionTime},\ \eqref{eq:yPrimeDistribution},\ 
	\eqref{eq:mseIntegration},\ 
	\eqref{eq:aoiIntegration}
	\end{align}
\end{subequations}
The achievability region of MSE and AoI is traversed through by iterating over a finite grid of $d$ and $\varepsilon$, see e.g. \cite{6924852}. The channel blocklength $n$ depends directly on these two variables. In order to enable a fast computation, closed-form expressions of the objectives as a function of $d$ and $\varepsilon$ are developed next for single-variate systems (the $k=1$ case).

\vspace{.1in}
 
\section{Closed-Form Expressions For $k=1$}

In this section, closed-form expressions of the time-average MSE and AoI are derived for fixed values of $d$ and $\varepsilon$, for the special case of having single-variate systems, i.e., $k=1$.

\subsection{Mean Square Error}

For $k=1$, matrices and vectors become scalars. Hence, the redefined variables $x(t)=\mathbf{x}(t)$, $\hat{x}(t)=\mathbf{\hat{x}}(t)$, $y(t)=\mathbf{y}(t)$, $q_u=\mathbf{Q}_u$, $q_w=\mathbf{Q}_w$, $q_x=\mathbf{Q}_x$, and $q(t)=\mathbf{Q}(t)$ are denoted in non-bold font for convenience. 
The individual components of the MSE are described in \eqref{eq:MnoChannelNoiseSolution} and \eqref{eq:MnoDelay} and can now be obtained as
\begin{align}
M^{\mathrm{D}}(t)&=q_u\int_{0}^{\tau}e^{2a(\tau-\mu)}d\mu=-\frac{q_u}{2a}+\frac{q_u}{2a}e^{2a\tau}, \\
 M^{\mathrm{C}}(t)&=\frac{q_xq_w}{q_x+q_w}e^{2a\tau}.
\end{align}
Thereby, from \eqref{eg:motionX} it can be obtained that $q_x=-\frac{q_u}{2a}$. When utilizing this, the packet-wise integral of the MSE stated in \eqref{eq:mseIntegration} can be derived as
\begin{align}
L(r,s,r')&=\int_{r}^{r+s+r'}\Xi+\Upsilon e^{2a\tau}d\tau\label{eq:mseIntegrationSISO1}\\
&=\Xi\left(s+r'\right)+\frac{\Upsilon}{2a}\left(e^{2a\left(r+s+r'\right)}-e^{2ar}\right),
\end{align}
where
\begin{align}
\Xi&=-\frac{q_u}{2a}\hspace{0.75cm}\text{and}\hspace{0.75cm}
\Upsilon=\frac{q_u}{2a}+\frac{q_xq_w}{q_x+q_w}.\nonumber
\end{align}
In order to obtain the MSE as described in \eqref{eq:limMFinite}, we need the expected value of the term $L(r,s,r')$ above. This can be derived using \eqref{eq:yPrimeDistribution} as follows:
\begin{align}
&\mathds{E}\left[L(r,s,r')\right]\nonumber\\&
=\sum_{m=0}^{\infty}\varepsilon^m(1-\varepsilon)\bigg(\Xi\left(s+(m+1)r\right)\nonumber\\&\hspace{2.5cm}+\frac{\Upsilon}{2a}\left(e^{2a\left((m+2)r+s\right)}-e^{2ar}\right)\bigg)\\
&=\sum_{m=0}^{\infty}m\varepsilon^m(1-\varepsilon)\Xi r\nonumber\\&\hspace{0.6cm}+\sum_{m=0}^{\infty}\varepsilon^m(1-\varepsilon)\left(\Xi\left(s+r\right)-\frac{\Upsilon}{2a} e^{2ar}\right)\nonumber\\&\hspace{0.6cm}+\sum_{m=0}^{\infty}e^{(\ln(\varepsilon)+2ar)m}(1-\varepsilon)\frac{\Upsilon}{2a} e^{2a(2r+s)}.
\end{align}
Simplifying the individual terms further leads to
\begin{align}
&\mathds{E}\left[L(r,s,r')\right]
\nonumber\\
&=\frac{\varepsilon}{1-\varepsilon}\Xi r +\Xi\left(s+r\right)-\frac{\Upsilon}{2a}e^{2ar}\nonumber\\&\hspace{0.9cm}+\frac{1-\varepsilon}{1-\varepsilon e^{2ar}}\frac{\Upsilon}{2a}e^{2a(2r+s)}\\
&=\Xi\left(\frac{1}{1-\varepsilon}r+s\right)+\frac{\Upsilon}{2a}\left(\frac{1-\varepsilon}{1-\varepsilon e^{2ar}}e^{2a(2r+s)}-e^{2ar}\right).
\end{align}
Finally, when inserting this into \eqref{eq:limMFinite}, the time-average MSE reduces to
\begin{align}
\mathrm{MSE}=\Xi+\frac{\Upsilon}{2a}\frac{\frac{1-\varepsilon}{1-\varepsilon e^{2ar}}e^{2a(2r+s)}-e^{2ar}}{\frac{1}{1-\varepsilon}r+s}.\label{eq:limMFinite2}
\end{align}
\subsection{Age of Information}
The time-average AoI is stated in \eqref{eq:aoiAvg}, which depends on $r$ and the expectation of $r'$. Whereas the former is equal to \eqref{eq:lengthToTransmissionTime}, the latter can be derived using \eqref{eq:yPrimeDistribution} as
\begin{align}
	\mathds{E}\left[r'\right]&=\sum_{m=0}^{\infty}\varepsilon^m(1-\varepsilon)(m+1)\mathds{E}\left[r\right]=\frac{1}{1-\varepsilon}r.\label{eq:yPrime}
\end{align}
Besides, \eqref{eq:aoiAvg} also depends on the second-order-moment $\mathds{E}\left[\left(r'\right)^2\right]$, which can be calculated as
\begin{align}
\mathds{E}\left[\left(r'\right)^2\right]&=\sum_{m=0}^{\infty}\mathds{E}\left[\left((m+1)y\right)^2\varepsilon^m(1-\varepsilon)\right]\nonumber\\
&=(1-\varepsilon)\mathds{E}\left[r^2\right]\frac{\varepsilon+1}{(1-\varepsilon)^3}=\frac{\varepsilon+1}{(1-\varepsilon)^2}r^2.\label{eq:yPrimeSecondMoment}
\end{align}
When utilizing \eqref{eq:yPrime} and \eqref{eq:yPrimeSecondMoment}, \eqref{eq:aoiAvg} can be expressed as
\begin{align}
\mathrm{AoI}&=\frac{\left(s+\frac{1}{1-\varepsilon}r\right)r+\frac{1}{2}\left(s^2+\frac{1}{1-\varepsilon}rs+\frac{\varepsilon+1}{\left(1-\varepsilon\right)^2}r^2\right)}{s+\frac{1}{1-\varepsilon}r}\nonumber\\
&=\frac{\frac{1}{2}(1-\varepsilon)s^2+(3-2\varepsilon)sr+\frac{3-\varepsilon}{2(1-\varepsilon)}r^2}{(1-\varepsilon)s+r}.\label{eq:aoiAvg2}
\end{align}


\section{Numerical Results}

In this section, we present some numerical results to further illustrate the results of the paper. We focus on the scalar case of $k=1$. For a given distortion level $d$ and violation probability $\varepsilon$, the channel blocklength $n$ and the transmission delay $r$ can be calculated, and hence the distribution of $r^\prime$ can be fully characterized. We consider a worst-case scenario of having $q_w$ exactly equal to $d$. The system is characterized by $a=-0.02$ and $q_u=1$. The SNR of the channel is $P=10$ and the waiting policy is specified by $s=0$ (zero-waiting).

\begin{figure}[t]
	\centering
	\includegraphics[scale=.89]{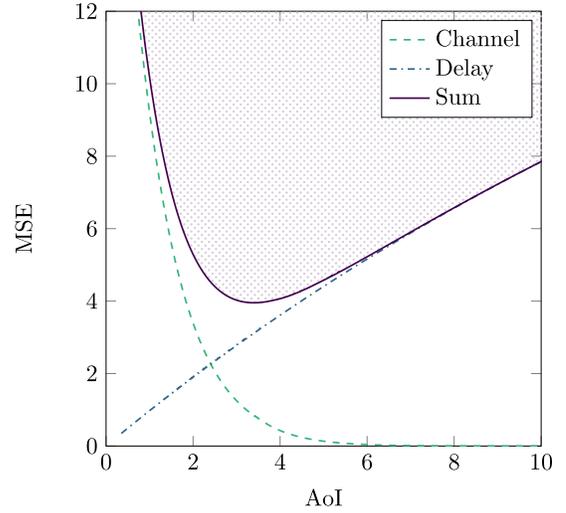}
	\vspace{-0.2cm}
	\caption{Achievable MSE together with its components of transmission delay and channel noise vs. AoI. Points constituting the dark-dotted region are achievable (MSE, AoI) pairs.}
	\label{fig:MseAndAoiComp}
\end{figure}

As stated before, the MSE consists of two parts: $\mathrm{MSE}^D$ (which we refer to as ``Delay'') and $\mathrm{MSE}^C$ (which we refer to as ``Channel''). The total MSE is given by their sum ($\mathrm{MSE}=\mathrm{MSE}^D+\mathrm{MSE}^C$) (which we refer to as ``Sum''). All components are shown in Fig.~\ref{fig:MseAndAoiComp}.

$\mathrm{MSE}^D$, in particular, describes the effects weakening the MSE when having a large AoI, and hence it can be seen in the figure that it is monotonically increasing with AoI. On the other hand, $\mathrm{MSE}^C$ describes the effects of distortion and noise, i.e., having reduced data resolution, on the MSE, which can be seen to be monotonically decreasing with the AoI in the figure. Their sum represents the total MSE and is shown to have a non-monotonic behavior with AoI. This sum curve represents the boundary of the achievable (MSE, AoI) region, where all values below the sum curve are not achievable.

The boundary shows that there exists an intrinsic relationship between MSE and AoI. Basically, precise information requires more transmission time whereas fast transmissions incur higher distortions. One can make either $\mathrm{MSE}^D$ or $\mathrm{MSE}^C$ sufficiently small, but not simultaneously. The boundary shows that neither small or large values of AoI are MSE-optimal. Instead, there exists an optimal point for AoI beyond which the MSE is not enhanced since larger delays are incurred, and before which the MSE is also not enhanced since larger distortions are incurred.

\section{Conclusion}

When monitoring physical or technical processes, there are different metrics describing the performance of the monitoring. In this paper, the trade-off between two of these metrics, namely the AoI and MSE, has been investigated. The MSE is mainly impacted by delays within the system and the transmission noise, whereas the AoI only depends on the system delays. Hence, both objectives show a similar behavior when the delays are large. However, if the system delays are small and the transmission noise is high, the MSE and AoI show the exact opposite behavior. Indeed, the AoI will be minimized when transmitting minimal data within each packet, whereas this would also lead to a large MSE. 

As an increased transmission time leads to changes of the process value to happen even before a data-packet is received, an upper limit of the per-packet information should not be exceeded. 
When exceeding this value, there will be neither a performance gain for the MSE, nor for the AoI. Instead, increasing the transmitted information too much will even lead to a reduction of both optimization variables. This illustrates that there is no need for reducing the distortion to an infinitesimally small value. Hence, the data accuracy and sampling quality can be reduced up to a certain limit without loss of performance. This approach allows having faster transmission times and a higher data frequency, which is an important factor for real-time applications.


\begin{thebibliography}{10}
\providecommand{\url}[1]{#1}
\csname url@samestyle\endcsname
\providecommand{\newblock}{\relax}
\providecommand{\bibinfo}[2]{#2}
\providecommand{\BIBentrySTDinterwordspacing}{\spaceskip=0pt\relax}
\providecommand{\BIBentryALTinterwordstretchfactor}{4}
\providecommand{\BIBentryALTinterwordspacing}{\spaceskip=\fontdimen2\font plus
\BIBentryALTinterwordstretchfactor\fontdimen3\font minus
  \fontdimen4\font\relax}
\providecommand{\BIBforeignlanguage}[2]{{%
\expandafter\ifx\csname l@#1\endcsname\relax
\typeout{** WARNING: IEEEtran.bst: No hyphenation pattern has been}%
\typeout{** loaded for the language `#1'. Using the pattern for}%
\typeout{** the default language instead.}%
\else
\language=\csname l@#1\endcsname
\fi
#2}}
\providecommand{\BIBdecl}{\relax}
\BIBdecl

\bibitem{5452208}
Y.~{Polyanskiy}, H.~V. {Poor}, and S.~{Verd{\'u}}, ``Channel coding rate in the
  finite blocklength regime,'' \emph{IEEE Transactions on Information Theory},
  vol.~56, no.~5, pp. 2307--2359, May 2010.

\bibitem{6408177}
V.~{Kostina} and S.~{Verd{\'u}}, ``Lossy joint source-channel coding in the
  finite blocklength regime,'' \emph{IEEE Transactions on Information Theory},
  vol.~59, no.~5, pp. 2545--2575, May 2013.

\bibitem{6195689}
S.~{Kaul}, R.~{Yates}, and M.~{Gruteser}, ``Real-time status: How often should
  one update?'' in \emph{2012 Proceedings IEEE INFOCOM}, March 2012, pp.
  2731--2735.

\bibitem{8006542}
Y.~{Sun}, Y.~{Polyanskiy}, and E.~{Uysal-Biyikoglu}, ``Remote estimation of the
  wiener process over a channel with random delay,'' in \emph{2017 IEEE
  International Symposium on Information Theory (ISIT)}, June 2017, pp.
  321--325.

\bibitem{DBLP:journals/corr/abs-1902-03552}
\BIBentryALTinterwordspacing
T.~Z. Ornee and Y.~Sun, ``Sampling for remote estimation through queues: Age of
  information and beyond,'' \emph{CoRR}, vol. abs/1902.03552, 2019. [Online].
  Available: \url{http://arxiv.org/abs/1902.03552}
\BIBentrySTDinterwordspacing

\bibitem{8006504}
E.~{Najm}, R.~{Yates}, and E.~{Soljanin}, ``Status updates through {M/G/1/1}
  queues with harq,'' in \emph{2017 IEEE International Symposium on Information
  Theory (ISIT)}, June 2017, pp. 131--135.

\bibitem{8445909}
H.~{Sac}, T.~{Bacinoglu}, E.~{Uysal-Biyikoglu}, and G.~{Durisi}, ``Age-optimal
  channel coding blocklength for an {M/G/1} queue with harq,'' in \emph{2018
  IEEE 19th International Workshop on Signal Processing Advances in Wireless
  Communications (SPAWC)}, June 2018, pp. 1--5.

\bibitem{8437671}
R.~{Devassy}, G.~{Durisi}, G.~C. {Ferrante}, O.~{Simeone}, and
  E.~{Uysal-Biyikoglu}, ``Delay and peak-age violation probability in
  short-packet transmissions,'' in \emph{2018 IEEE International Symposium on
  Information Theory (ISIT)}, June 2018, pp. 2471--2475.

\bibitem{8849713}
E.~{Najm}, E.~{Telatar}, and R.~{Nasser}, ``Optimal age over erasure
  channels,'' in \emph{2019 IEEE International Symposium on Information Theory
  (ISIT)}, July 2019, pp. 335--339.

\bibitem{7925903}
P.~{Parag}, A.~{Taghavi}, and J.~{Chamberland}, ``On real-time status updates
  over symbol erasure channels,'' in \emph{2017 IEEE Wireless Communications
  and Networking Conference (WCNC)}, March 2017, pp. 1--6.

\bibitem{8006541}
R.~D. {Yates}, E.~{Najm}, E.~{Soljanin}, and J.~{Zhong}, ``Timely updates over
  an erasure channel,'' in \emph{2017 IEEE International Symposium on
  Information Theory (ISIT)}, June 2017, pp. 316--320.

\bibitem{8377368}
E.~T. {Ceran}, D.~{G{\"u}nd{\"u}z}, and A.~{Gy{\"o}rgy}, ``Average age of
  information with hybrid arq under a resource constraint,'' in \emph{2018 IEEE
  Wireless Communications and Networking Conference (WCNC)}, April 2018, pp.
  1--6.

\bibitem{DBLP:journals/corr/abs-1905-03238}
\BIBentryALTinterwordspacing
A.~Arafa, K.~Banawan, K.~G. Seddik, and H.~V. Poor, ``On timely channel coding
  with hybrid {ARQ},'' \emph{CoRR}, vol. abs/1905.03238, 2019. [Online].
  Available: \url{http://arxiv.org/abs/1905.03238}
\BIBentrySTDinterwordspacing

\bibitem{DBLP:journals/corr/abs-1903-12472}
\BIBentryALTinterwordspacing
K.~Huang, W.~Liu, M.~Shirvanimoghaddam, Y.~Li, and B.~Vucetic, ``Real-time
  remote estimation with hybrid {ARQ} in wireless networked control,''
  \emph{CoRR}, vol. abs/1903.12472, 2019. [Online]. Available:
  \url{http://arxiv.org/abs/1903.12472}
\BIBentrySTDinterwordspacing

\bibitem{8761710}
K.~{Huang}, W.~{Liu}, Y.~{Li}, and B.~{Vucetic}, ``To retransmit or not:
  Real-time remote estimation in wireless networked control,'' in \emph{ICC
  2019 - 2019 IEEE International Conference on Communications (ICC)}, May 2019,
  pp. 1--7.

\bibitem{8619768}
J.~{Yun}, C.~{Joo}, and A.~{Eryilmaz}, ``Optimal real-time monitoring of an
  information source under communication costs,'' in \emph{2018 IEEE Conference
  on Decision and Control (CDC)}, Dec 2018, pp. 4767--4772.

\bibitem{8814627}
A.~{Mitra}, J.~A. {Richards}, S.~{Bagchi}, and S.~{Sundaram}, ``Finite-time
  distributed state estimation over time-varying graphs: Exploiting the
  age-of-information,'' in \emph{2019 American Control Conference (ACC)}, July
  2019, pp. 4006--4011.

\bibitem{7282683}
J.~{Chakravorty} and A.~{Mahajan}, ``Distortion-transmission trade-off in
  real-time transmission of gauss-markov sources,'' in \emph{2015 IEEE
  International Symposium on Information Theory (ISIT)}, June 2015, pp.
  1387--1391.

\bibitem{8988940}
M.~{Bastopcu} and S.~{Ulukus}, ``Age of information for updates with
  distortion,'' in \emph{2019 IEEE Information Theory Workshop (ITW)}, Aug
  2019, pp. 1--5.

\bibitem{7076051}
L.~{Rambault}, E.~{Etien}, S.~{Cauet}, G.~{Champenois}, and O.~{Bachelier},
  ``Linearization by using two methods: Exact input-output linearization and
  sliding mode control,'' in \emph{2001 European Control Conference (ECC)},
  Sep. 2001, pp. 1040--1043.

\bibitem{Astrom04feedbacksystems}
K.~J. {\AA}str{\"o}m and R.~M. Murray, ``Feedback systems: An introduction for
  scientists and engineers,'' Tech. Rep., 2004.

\bibitem{Cover:2006:EIT:1146355}
T.~M. Cover and J.~A. Thomas, \emph{Elements of Information Theory (Wiley
  Series in Telecommunications and Signal Processing)}.\hskip 1em plus 0.5em
  minus 0.4em\relax New York, NY, USA: Wiley-Interscience, 2006.

\bibitem{6145679}
V.~{Kostina} and S.~{Verd{\'u}}, ``Fixed-length lossy compression in the finite
  blocklength regime,'' \emph{IEEE Transactions on Information Theory},
  vol.~58, no.~6, pp. 3309--3338, June 2012.

\bibitem{e20060399}
\BIBentryALTinterwordspacing
J.~Guti{\'e}rrez-Guti{\'e}rrez, M.~Z{\'a}rraga-Rodr{\'\i}guez, F.~M.
  Villar-Rosety, and X.~Insausti, ``Rate-distortion function upper bounds for
  {Gaussian} vectors and their applications in coding ar sources,''
  \emph{Entropy}, vol.~20, no.~6, 2018. [Online]. Available:
  \url{https://www.mdpi.com/1099-4300/20/6/399}
\BIBentrySTDinterwordspacing

\bibitem{8000687}
Y.~{Sun}, E.~{Uysal-Biyikoglu}, R.~D. {Yates}, C.~E. {Koksal}, and N.~B.
  {Shroff}, ``Update or wait: How to keep your data fresh,'' \emph{IEEE
  Transactions on Information Theory}, vol.~63, no.~11, pp. 7492--7508, Nov
  2017.

\bibitem{6924852}
E.~{Bjornson}, E.~A. {Jorswieck}, M.~{Debbah}, and B.~{Ottersten},
  ``Multiobjective signal processing optimization: The way to balance
  conflicting metrics in {5G} systems,'' \emph{IEEE Signal Processing
  Magazine}, vol.~31, no.~6, pp. 14--23, Nov 2014.

\end{thebibliography}


\end{document}